# Quantization of mental stress using various physiological markers

**Apoorvagiri[1*], Mandya Sannegowda Nagananda[2]**

[1]Biomedical signal processing and instrumentation,

RV College of Engineering,

Bangalore, Karnataka, India

[2] Instrumentation and Technology,

R V College of Engineering,

Bangalore - 560 059, Karnataka, India

*Corresponding Author:

Apoorvagiri:

Apoorvagiri@gmail.com

**Abstract**

The aim of this study is to quantize mental stress by integrating different physiological markers like reaction time, photoplethysmograph (PPG), heart rate variability (HRV) and subjective markers like questionnaire. The study included 10 subjects of age between 22 and 26 years. Study materials included the results of PSS questionnaire, simple reaction time, PPG data, and HRV data during a stress inducing stroop test. The study suggests that mental stress can be quantized when stress is induced acquisitively and more accurate quantification of stress can be achieved by integrating many physiological parameters.

**Introduction**

Mental stress is considered as physiological response to the mental, emotional, or physical challenges. Most of the times mental stress provokes body's "fight or flight" response called as acute stress response. But prolonged or chronic stress can affect numerous physiological functions, such as growth, immune system, metabolism, reproduction and circulation [Charmandari, Tsigos, Chrousos, 2005]. Human body maintains homeostasis, which is frequently confronted by various internal and external factors called stressors they can be real or perceived, physical or mental. These stressors will lead to instant stimulation of autonomic nervous system (ANS), and also resulting in increased and decreased activity of sympathetic (SNS) and parasympathetic nervous system (PNS) [Pignatelli Magalhaes, Magalhaes C, 1998]. Since the change activity of the bodily systems are needed for "fight or flight" response in short term, but if the response is prolonged leading to delayed or no stress response will lead to serious health disorders [McEwen, 1998 ].

Continuous monitoring and measurement of mental stress levels are essentially required for assessing and managing routine mental stress. Many physiological markers are used including galvanic skin response, heart beat patterns (HRV), respiration, Finger pulse rate (PPG). Individuals can closely track changes in vital sign using modern wearable devices but measuring vital signs during routine activity is prone to irresistible noise. The same process can be achieved in laboratory.

Chong Zhang, Xiaolin Yu . [2010] proposed the effects of long term mental arithmetic task on psychology are investigated by subjective self-reporting measures and action performance test. Based on electroencephalogram (EEG) and heart rate variability (HRV), the impacts of prolonged cognitive activity on central nervous system and autonomic nervous system are observed and analyzed. Kwang Shin Park et al. [2011] obtained EEG (electroencephalogram) data from 34 healthy subjects while they were watching emotion-inducing videos and they also developed a real-time emotion monitoring system based on the resulting data. Maurizio Mauri et al. [2010] presented a preliminary quantitative study aimed at developing an optimal standard protocol for automatic classification of specific affective states as related to human-computer interactions. Yuan-Pin Lin et al. [2010] applied machine learning algorithms to categorize EEG dynamics according to subject self-reported emotional states during music listening. Seizi Nishifuji. [2011] investigated response of electroencephalogram (EEG) to aerobic exercise with low intensity after performing mental task with listening to acoustic stimuli in order to measure a recovery effect of the acute exercise on the EEG. Christos et al [2010] study proposes a methodology for the robust classification of neurophysiological data into four emotional states collected during passive viewing of emotional evocative pictures selected from the International Affective Picture System. In literature researchers have considered single parameters to assess stress; considering questionnaire and reaction time can increase substantially accuracy in measuring mental stress. This motivates us to consider more than one parameter to measure and assess stress.

The aim of this study is to establish mental stress assessment protocol by combining different physiological parameters. Questionnaire was used as a qualitative assessment and Reaction time, HRV analysis and PRV analysis are used as a quantitative assessment of mental stress.

**Methodology**

The methodology consists of four principal steps

1.  Taking questionnaire from the subject based on Perceived stress scale

2.  Recording Reaction time data when subject takes reaction time test

3.  Acquiring ECG and PPG from the subject and extracting features

4. Classification of subject into low, medium and highly stress using a neural network

**Methodology for stress assessment**

The stress assessment is done through three principle parameters they are

- Questionnaire
- Reaction time test
- Using physiological signals

**Questionnaire:** Questionnaire is frequently used as a measure of mental well being with those people with values below a certain threshold regarded as suffering from mental stress. Percieved Stress Scale Questionnaire was developed to assess the impact of emotional, financial and academic stressors.

**Reaction time test**: Reaction time is a measure of how quickly an organism can respond to a particular stimulus. Reaction time has been widely studied[Jaworski, Janusz, et al(2013a, 2013b)], Apoorvagiri, Nagananda(2013)], as its practical implications may be of great consequence, e.g. a slower than normal reaction time while driving can have grave results. Many factors have been shown to affect reaction times, including age, gender, physical fitness, fatigue, distraction, alcohol, personality type, and whether the stimulus is auditory or visual. Refers to how long it takes a person to respond to a given stimulus. Since reaction time and stress are related we designed a reaction time test from which stress can be assessed.

**Using Physiological signals:** Although stress has a psychological origin, it affects several physiological processes like Heart Rate, Blood Pressure, Skin Conductance, Electroencephalogram, Reaction Time, Saliva Amylase, etc. Here we acquired ECG and PPG for stress assessment.

**Subjective Assessment:**

Questionnaire and reaction time test were used as protocol for subjective assessment. Figure 1 explains the protocol.

**Stroop test**

One's mental stress depends on one's knowledge and experience related to the problem as well as many other cognitive parameters. Given the same design problem, different people will have different brain activities which correspond to different EEG wave patterns. Therefore, it is indispensable to define a baseline in quantifying the mental stresses from different people. Stroop test is used to achieve this goal and this is the focus of this project. Stroop test, a color naming task, is a classical paradigm in neurophysiologic assessment of mental fitness. The Stroop test is a demonstration of interference in the reaction of the task. In our experiment, the Stroop test is designed as a computer game in which a subject is presented a color name, referred as stimulus word. The stimulus word is displayed in a color which is the same as or different from that it refers to. The subject has to select the answer corresponding to the color of the word. For example, given a GREEN word in BLUE color, the subject has to select the word BLUE in the answer list. Our Stroop test contains five colors: RED, BLUE, YELLOW, PURPLE and GREEN. Stroop test interface and protocol are shown in figure 2 and 3 respectively.

**Objective Assessment**

Figure 4 clearly represents objective assessment protocol where PPG and ECG are used to assess mental stress.

**ECG data acquisition:**

An ECG signal is acquired after the subject performs stroop test using National instruments modules. Consent form was signed by all the subjects. Figure 5 represents the block diagram of data acquisition using National instruments module. Three leads electrodes are placed on the subject from whom an ECG signal has to be acquired. The signals from the electrodes are amplified using a bioamplifier which comes along with National Instruments. The output of the bioamplifier is connected to the multichannel 6009 DAQ card, to acquire raw ECG signals from the output terminal of ECG recorders. The sampling rate is typically set to 125 Hz or 250 Hz. The acquired ECG signals can be stored in NI TDMS file type for offline analysis. DAQ is connected to the computer using an USB port for further processing of the signal. DAQ has to be configured inside LabVIEW environment.

**PPG data acquisition:**

The PPG signal is acquired after performing stroop test using Skrip Electronics modules. Consent form was signed by all the subjects. Figure 6 represents the block diagram of data acquisition using Skrip Electronics module. A reflectance type IR LED sensor is placed on the subject fingertip from whom a PPG signal has to be acquired. The signals from the electrodes are amplified using a bioamplifier which comes along with Skrip Electronics modules. The output of the bioamplifier is connected to the multichannel 6009 DAQ card, to acquire raw PPG signals from the output terminal of PPG recorders. The sampling rate is typically set to 256 Hz. The acquired PPG signals can be stored in NI TDMS file type for offline analysis. DAQ is connected to the computer using an USB port for further processing of the signal. DAQ has to be configured inside Lab VIEW environment.

For analysis of HRV data using this neural network tool is done to classify subjects into low, medium and highly stressed. The data base of acquired signal is done in an MS-Office Excel sheet with distinct features. Now, data is processed and trained to select appropriate network. After training the network we can give input HRV data's to classify the given data to check the condition of given subject.

Flow algorithm explaining ECG and PPG extraction and processing is shown in figure 7.

**Results and discussion**

**The Questionnaire and Reaction time test**

The below table shows PSS-5 scores and mean and standard deviation of reaction times for 100 and 300 words(figure 8) given in the stroop test. Table 1 showed detailed tabulation of results of PSS-5 scores and stroop test results.

The negative Pearson Correlation of -0.119 between PSS-5 Score and 100 words stroop test mean and Positive Pearson Correlation of 0.111 between PSS-5 Score and 300 words Stroop test mean. Hence 300 word stroop test is more effective to assess stress and further future research is pursued for 500 words and 1000 words correlation to efficiently assess stress.

**HRV and PRV extraction**

Heart rate variability (figure 9) and Pulse rate variability (figure 10) is acquired after subject performs stress inducing test. The entropy (randomness) is increased when subject is stressed. HRV and PRV entropy are both high in stressed situation.

**Tabulated training data**

The Pearson correlation between sample entropy of each subject and PSS score is found to be -0.943. Since low PSS score indicates high stress, correlation value proves that as stress increases sample entropy also increases.

**Classification using neural network:**

For analysis of HRV data using this neural network tool is done to classify data as low, medium and highly stressed subjects. The data base of acquired signal is done in an MS-Office Excel sheet with distinct features (table 2). Now, data is processed and trained to select appropriate network. After training the network we can give input HRV data's to classify the given data to check the condition of given subject.

Alyuda NeuroIntelligence tool is used to analyze and classify data to its respective stress condition of subject. This work includes a data collection of 10 subjects of different stress conditions; the acquired ECG data using LabVIEW tools are processed in order to calculate inter beat intervals of ECG data. This IBI data is used to extract the HRV parameters, which are imported in NeuroIntelligence tool to classify it as low, medium and highly stressed subject using appropriate neural network. Figure 11 clearly explains how NeuroIntelligence tool is used to quantify stress.

**Conclusion**

Stress has been a growing issue in all professions, there is an urgent need for efficient stress assessment methods in order to solve psycho-physiological problems. This could be possible by integrating many biomarkers in an integrated system and designing a better stress measuring device might help to provide solutions for stress disorders.

Efficient assessment of Mental Stress is possible by combining Questionnaire, Reaction time test, ECG and PPG. 11% negative correlation is achieved w.r.t PSS score and reaction time and 94.3% negative correlation is observed w.r.t sample entropy and PSS score The Correlation coefficient can be increased by combining more physiological parameters. There is a need for low cost embedded system for stress assessment.

Figures:

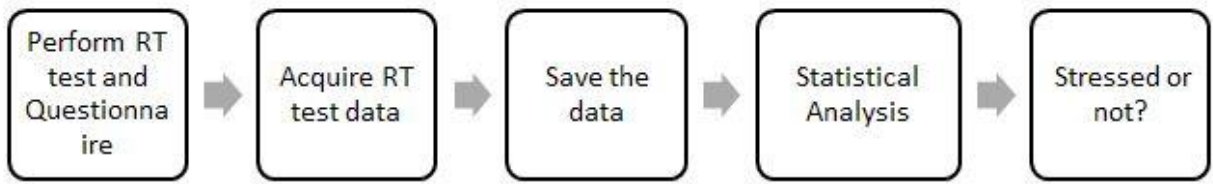

**Figure 1 Experimental protocol for questionnaire and reaction time test**

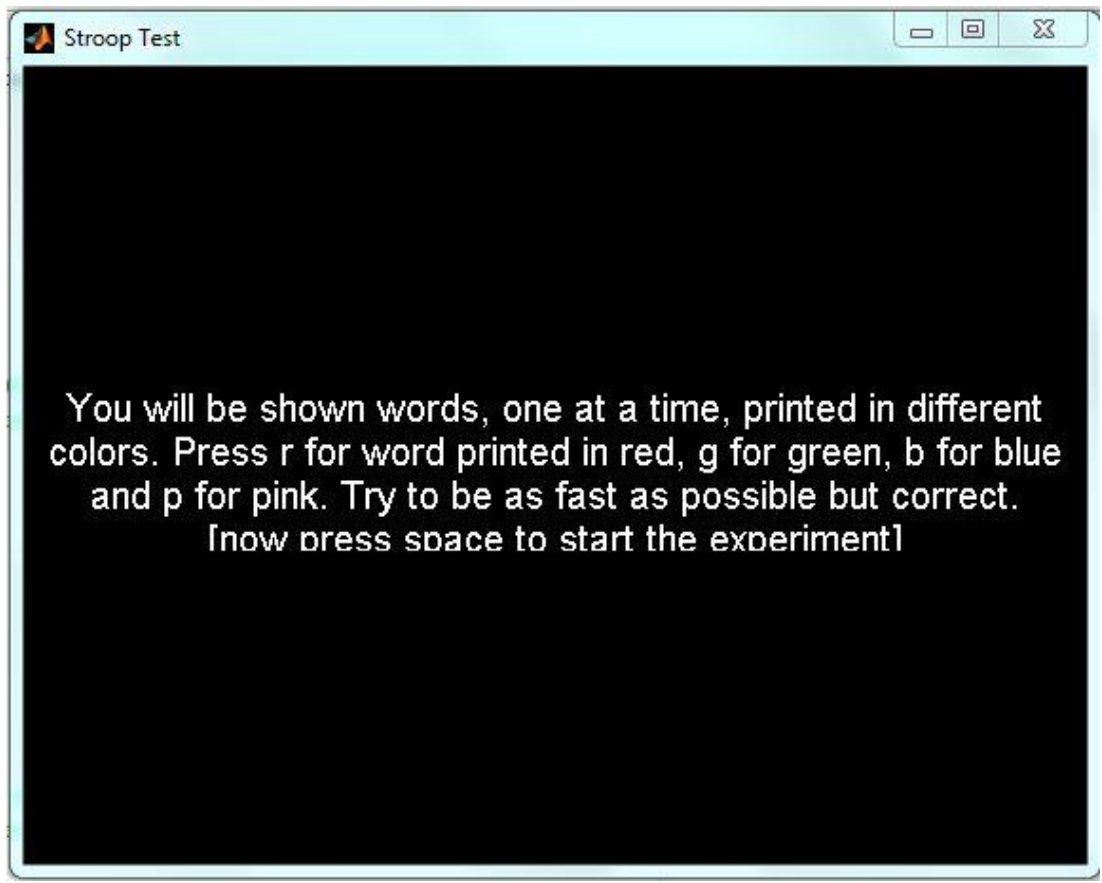

**Figure 2: Stroop test interface displaying instructions to subjects to start the stress inducing task.**

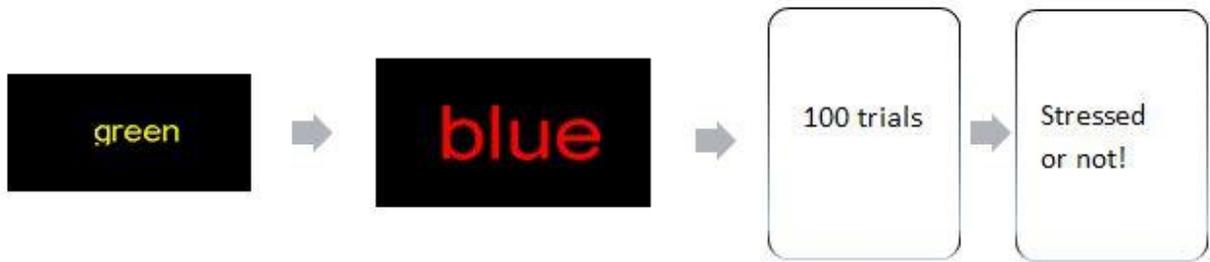

**Figure 3 Explaining Stroop test protocol where in which subjects were shown different color names colored in different colors the subject has to select the right color on the word.**

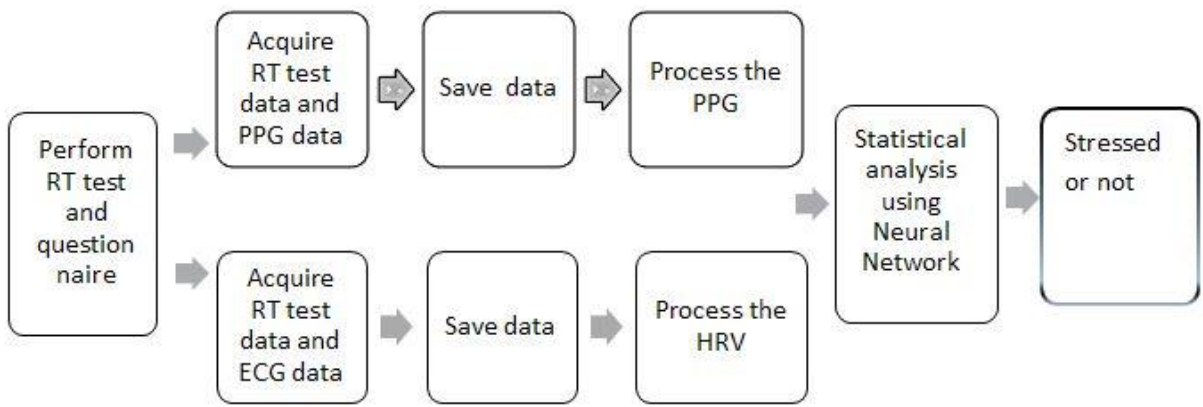

**Figure 4 Experimental protocol for ECG and PPG which is acquired after subject performs stress inducing task**

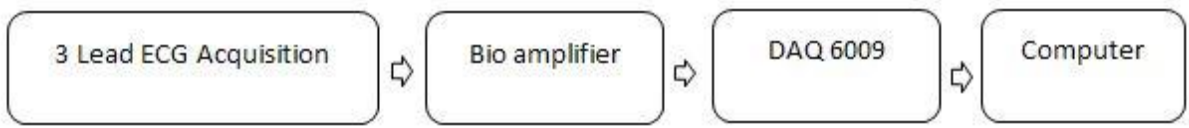

**Figure 5: Block diagram of data acquisition using National instruments module**

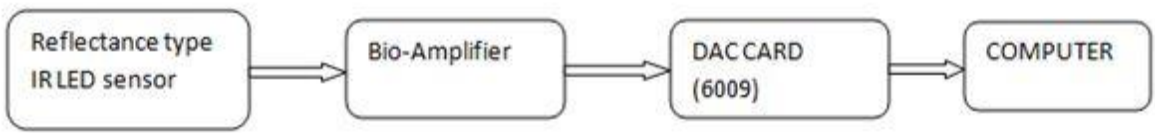

**Figure 6 Block diagram of data acquisition using skrip electronics modules.**

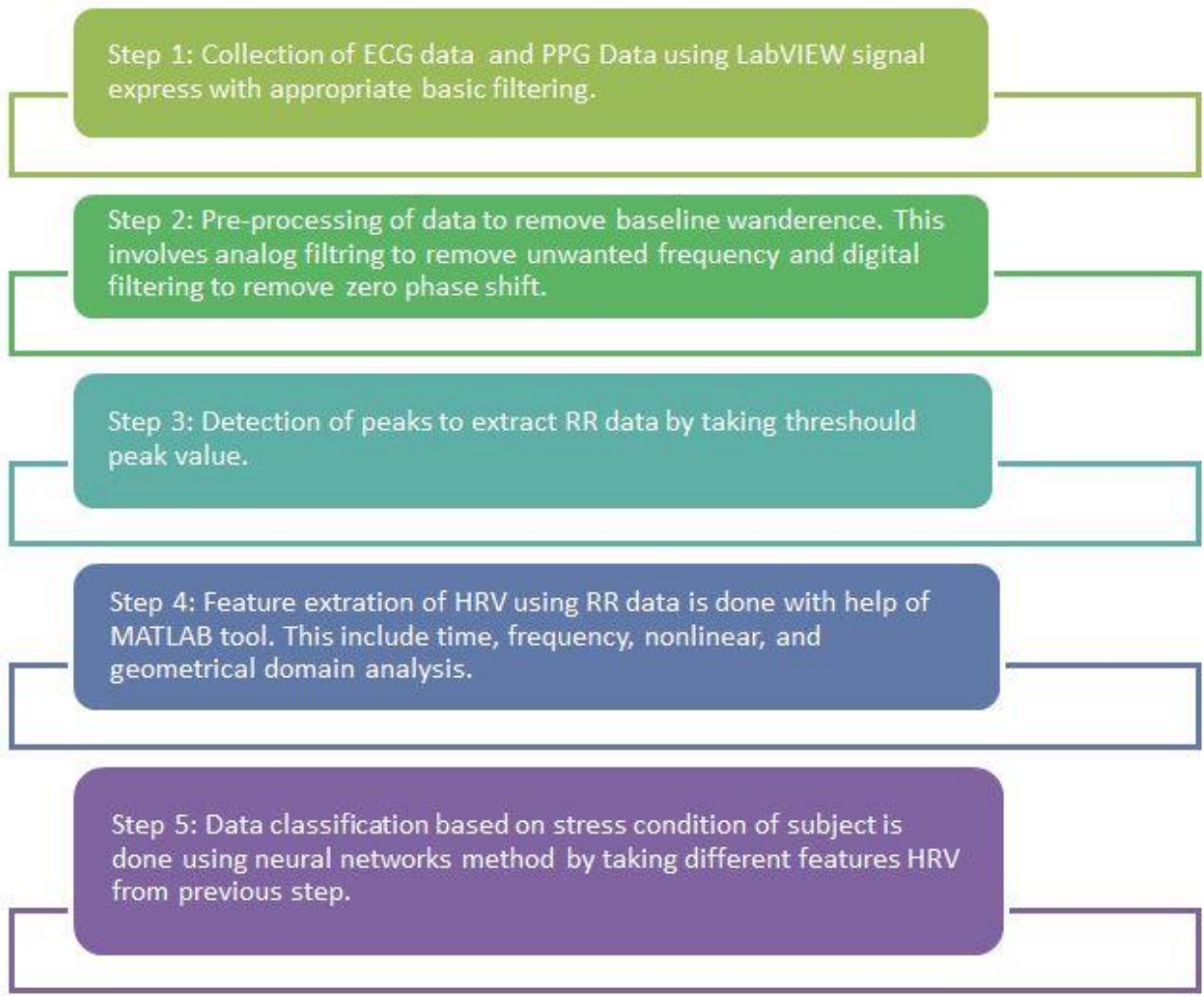

**Figure 7: Algorithm to assess ANS using HRV data.**

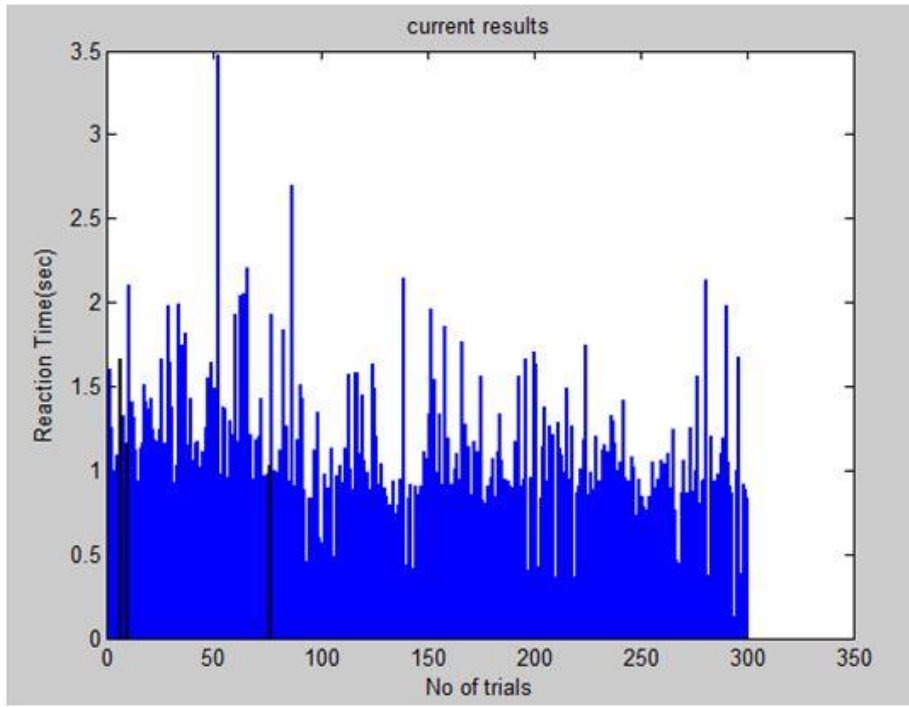

**Figure 8 Reaction time(in seconds) test result for 300 words**

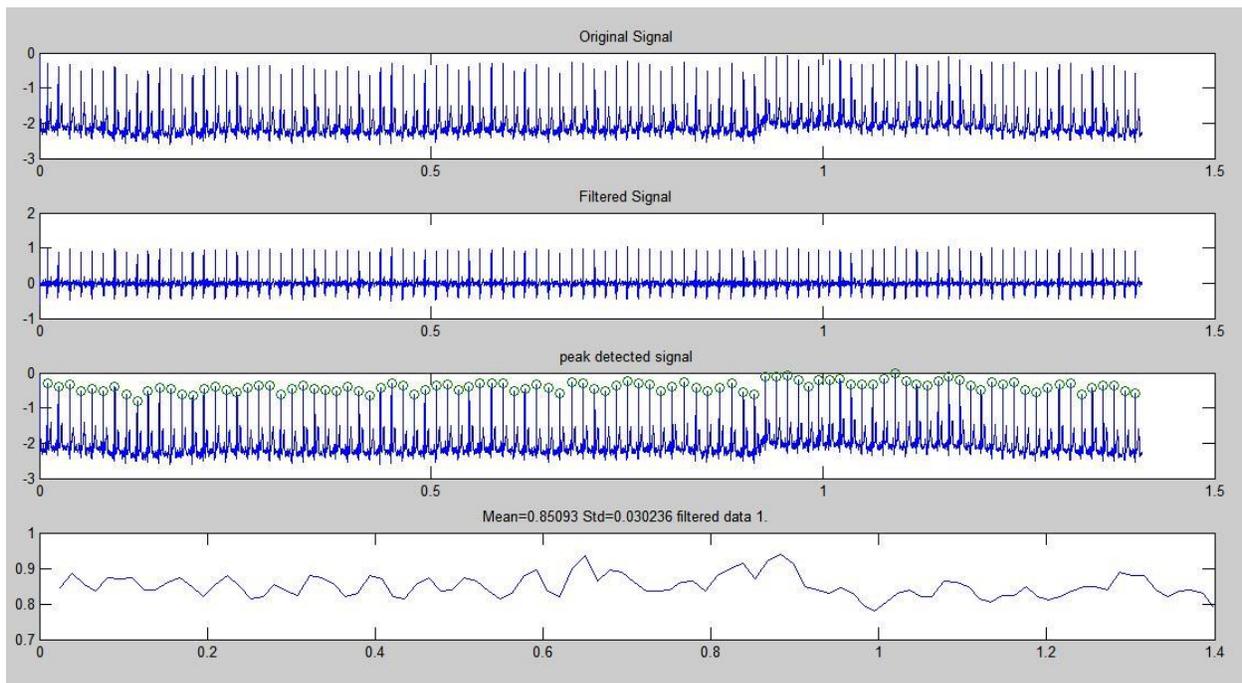

**Figure 9 Peak detection and HRV extraction, the final plot is heart rate variability whose change is instantaneous since the patient will be stressed after the task.**

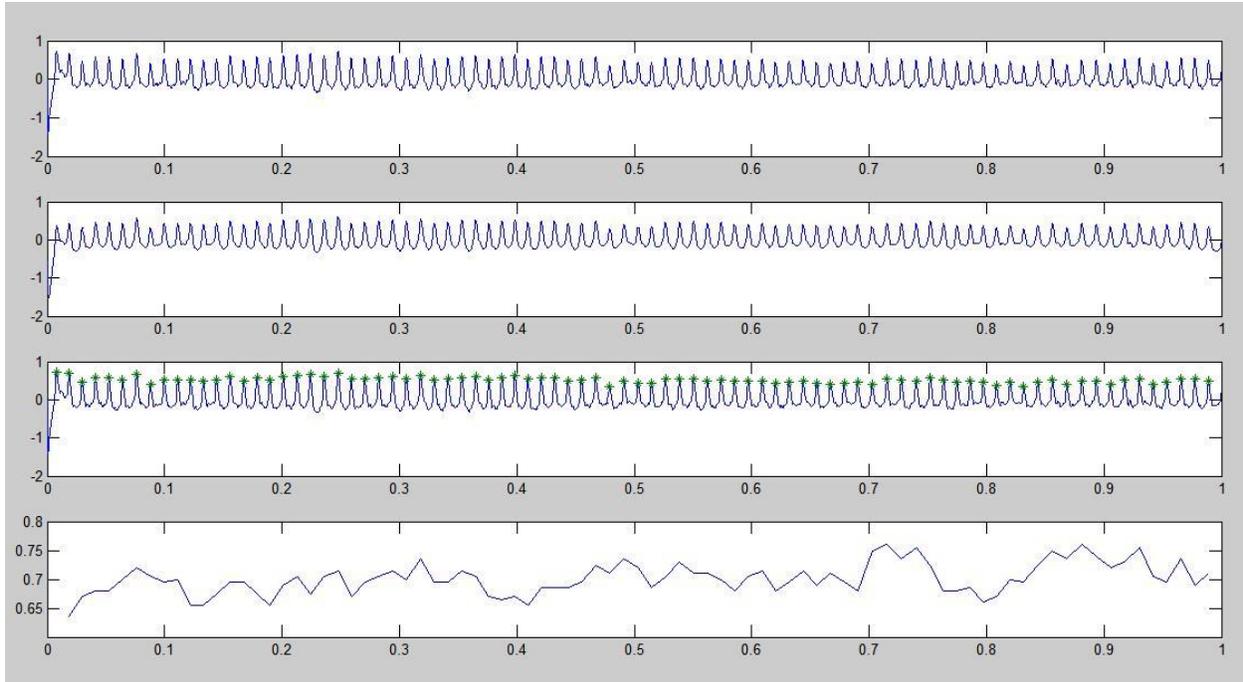

**Figure 10 Peak detection and PRV extraction, the final plot is pulse rate variability whose change is instantaneous since the patient will be stressed after the task.**

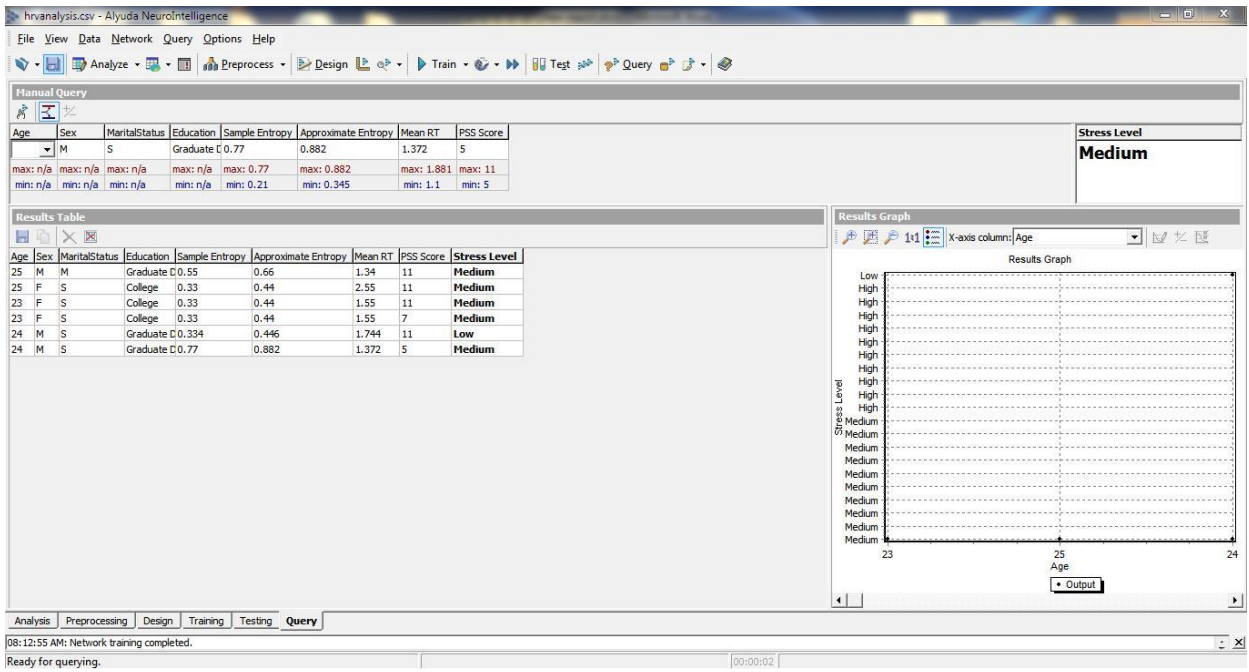

**Figure 11: A snapshot of tested result for given input data. Before giving input data the training data is used to train the neural network.**

**Tables**

| Subject | PSS-5 Score | 100 words Stroop test | | 300 words Stroop test | |
|---|---|---|---|---|---|
| | | Mean(sec) | S.D | Mean(sec) | S.D |
| Subject 1 | 7 | 1.565 | 0.673 | 1.881 | 1.205 |
| Subject 2 | 6 | 2.045 | 1.575 | 1.555 | 1.132 |
| Subject 3 | 9 | 1.124 | 0.321 | 1.158 | 0.384 |
| Subject 4 | 5 | 1.447 | 0.496 | 1.372 | 0.491 |
| Subject 5 | 11 | 1.586 | 0.756 | 1.744 | 1.136 |
| Subject 6 | 9 | 1.271 | 0.450 | 1.100 | 0.385 |
| Subject 7 | 7 | 1.454 | 0.706 | 1.563 | 0.924 |
| Subject 8 | 9 | 1.526 | 0.581 | 1.760 | 1.225 |
| Subject 9 | 11 | 1.862 | 1.086 | 1.881 | 1.205 |
| Subject 10 | 10 | 1.337 | 0.998 | 1.243 | 0.752 |

Table 1: Results showing PSS-5 Scores(least is more stressed) and Stroop test results for 100 words and 300 words stroop test.

| Subject No | Age | Sex | Marital Status | Education | Sample Entropy | Approximate Entropy | Mean RT(minutes) | PSS Score | Stress Level |
|---|---|---|---|---|---|---|---|---|---|
| Subject 1 | 23 | M | S | Graduate Degree | 0.581 | 0.682 | 1.881 | 7 | Medium |
| Subject 2 | 23 | M | S | Graduate Degree | 0.68 | 0.799 | 1.555 | 6 | Medium |
| Subject 3 | 25 | M | M | College | 0.45 | 0.564 | 1.158 | 9 | Medium |
| Subject 4 | 24 | F | S | Graduate Degree | 0.77 | 0.882 | 1.372 | 5 | High |
| Subject 5 | 23 | M | S | Graduate Degree | 0.334 | 0.446 | 1.744 | 11 | Low |

| Subject 6 | 24 | M | S | Graduate Degree | 0.49 | 0.594 | 1.1 | 9 | Medium |
| Subject 7 | 24 | M | S | Graduate Degree | 0.62 | 0.643 | 1.563 | 7 | Medium |
| Subject 8 | 23 | F | S | Graduate Degree | 0.59 | 0.66 | 1.76 | 9 | Medium |
| Subject 9 | 23 | F | S | Graduate Degree | 0.21 | 0.345 | 1.881 | 11 | Low |
| Subject 10 | 24 | M | S | Graduate Degree | 0.39 | 0.444 | 1.243 | 10 | Low |

Table 2: A snapshot of tabulated training data's (total 10 subjects)